\documentclass[superscriptaddress,twocolumn,showpacs,pra,floatfix]{revtex4}

\usepackage{epsfig}
\usepackage{color}
\usepackage{tabularx}
\usepackage{epsfig}
\usepackage{amsmath}
\usepackage{amssymb}
\usepackage{graphicx}
\usepackage{wasysym}
\usepackage{dcolumn}
\usepackage{bm}
\usepackage{psfrag}
\usepackage{times}

\begin{document}

\title{Lack of a thermodynamic finite-temperature spin-glass phase in
the two-dimensional randomly-coupled ferromagnet}

\author{Zheng Zhu}
\affiliation{Department of Physics and Astronomy, Texas A$\&M$ University,
College Station, Texas 77843-4242, USA}

\author{Andrew J.~Ochoa}
\affiliation{Department of Physics and Astronomy, Texas A$\&M$ University,
College Station, Texas 77843-4242, USA}

\author{Helmut G. Katzgraber}
\affiliation{Department of Physics and Astronomy, Texas A$\&M$ University,
College Station, Texas 77843-4242, USA}
\affiliation{1QB Information Technologies (1QBit), Vancouver, British
Columbia, Canada V6B 4W4}
\affiliation{Santa Fe Institute, 1399 Hyde Park Road, Santa Fe, New Mexico 
87501, USA}

\begin{abstract}

The search for problems where quantum adiabatic optimization might excel
over classical optimization techniques has sparked a recent interest in
inducing a finite-temperature spin-glass transition in quasi-planar
topologies. We have performed large-scale finite-temperature Monte Carlo
simulations of a two-dimensional square-lattice bimodal spin glass with
next-nearest ferromagnetic interactions claimed to exhibit a
finite-temperature spin-glass state for a particular relative strength
of the next-nearest to nearest interactions [Phys.~Rev.~Lett.~{\bf 76},
4616 (1996)]. Our results show that the system is in a paramagnetic
state in the thermodynamic limit, despite zero-temperature simulations
[Phys.~Rev.~B {\bf 63}, 094423 (2001)] suggesting the existence of a
finite-temperature spin-glass transition. Therefore, deducing the
finite-temperature behavior from zero-temperature simulations can be
dangerous when corrections to scaling are large.

\end{abstract}

\maketitle

\section{Introduction}
\label{sec:intro}

The advent of analog quantum annealing machines
\cite{nielsen:00,nishimori:01,finnila:94,kadowaki:98,brooke:99,farhi:00,roland:02,santoro:02,das:05,santoro:06,lidar:08,das:08,morita:08,mukherjee:15}
and, in particular, the D-Wave Inc.~\cite{comment:d-wave} D-Wave 2X
quantum annealer has sparked a new interest in the study of
(quasi-) planar Ising spin glasses
\cite{mezard:84,binder:86,young:98,stein:13} with finite-temperature
transitions.  While there have been multiple attempts to discern if the
D-Wave quantum annealers display an advantage over conventional
technologies
\cite{dickson:13,pudenz:13,smith:13,boixo:13a,albash:15a,ronnow:14a,katzgraber:14,lanting:14,santra:14,shin:14,boixo:14,albash:15,albash:15a,katzgraber:15,martin-mayor:15a,pudenz:15,hen:15a,venturelli:15a,vinci:15,zhu:16},
to date there are only few ``success stories''
\cite{katzgraber:15,denchev:16} where the analog quantum optimizers show
an advantage over current conventional silicon-based computers.  Recent
results \cite{katzgraber:14,katzgraber:15} suggest that problems with a
more complex energy landscape are needed to discern if quantum annealers
can outperform current digital computers. In particular, the search for
salient features in the energy landscape \cite{katzgraber:15}, the
careful construction of problems with particular features
\cite{katzgraber:15,katzgraber:15,denchev:16,zhu:16}, as well as the
attempt to induce a finite-temperature spin-glass transition for
lattices restricted to the quasi-two-dimensional topologies of the
quantum chips \cite{bunyk:14} have gained considerable attention. The
quest for a finite-temperature spin-glass transition in
quasi-two-dimensional topologies stems from the interest in creating an
energy landscape that becomes more complex and rugged already at finite
temperatures, such that thermal (sequential) simulated annealing
\cite{kirkpatrick:83} has a harder time in determining the optimal
solution to an Ising-spin-glass-like optimization problem. On the other
hand, quantum annealing should, in principle, be able to tunnel through
barriers if these are thin enough. We emphasize that the comparison
between simulated annealing --- a well-known poor optimizer --- and
quantum annealing is based on the fact that both methods are sequential
in nature. Comparisons to state-of-the-art optimization techniques
\cite{mandra:16b} have been performed and shed a more complete light on
the current situation.

Here we want to study the thermodynamic properties of a model proposed
by Lemke and Campbell \cite{lemke:96}---later analyzed in much detail in
Refs.~\cite{parisi:98,lemke:99,hartmann:01d}---that might have the
desired finite-temperature spin-glass transition and, most importantly,
be of a mostly planar topology that can easily be constructed with
current superconducting flux qubits. Our results show that,
unfortunately, for large enough system sizes the model is in a
paramagnetic phase at finite temperatures for a parameter range where it
is predicted to be a spin glass. We do note that this would have been
surprising, because there is solid evidence that the lower critical
dimensions of spin glasses is believed to be between two and three space
dimensions \cite{bray:84,hartmann:01,boettcher:05d}---a value below
which any phase transition to a spin-glass state only occurs at zero
temperature.

The paper is structured as follows: In Sec.~\ref{sec:model} we describe
the model and numerical details, as well as the current understanding of
its properties, followed by results and concluding remarks 
in Sec.~\ref{sec:results}.

\begin{figure*}[!t]
\center
\includegraphics[width=0.96\columnwidth]{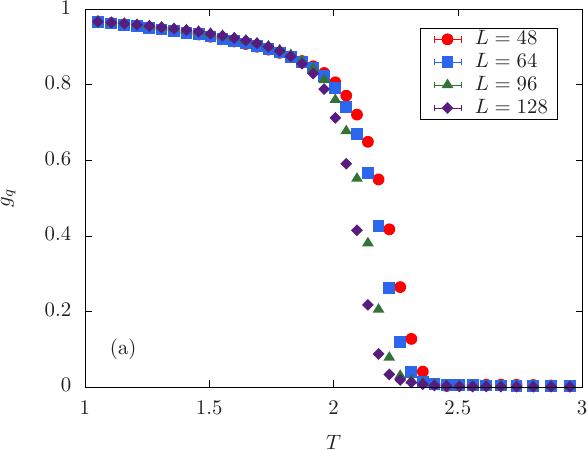}
\includegraphics[width=\columnwidth]{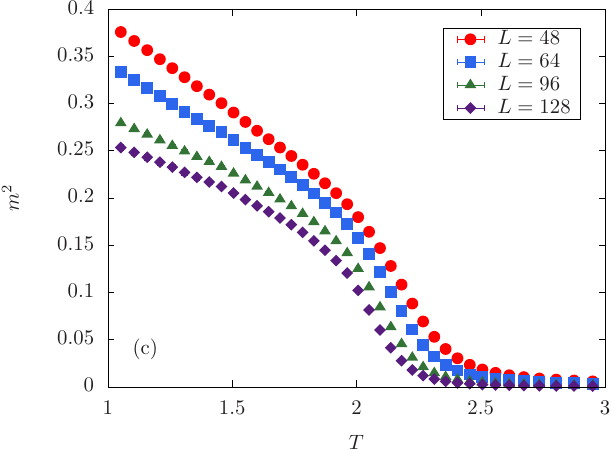}

\includegraphics[width=0.96\columnwidth]{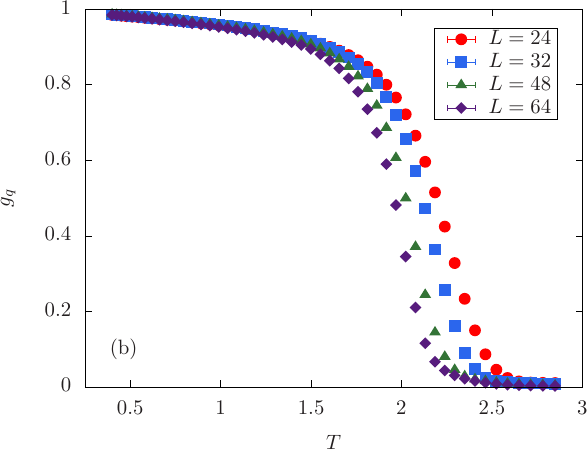}
\includegraphics[width=\columnwidth]{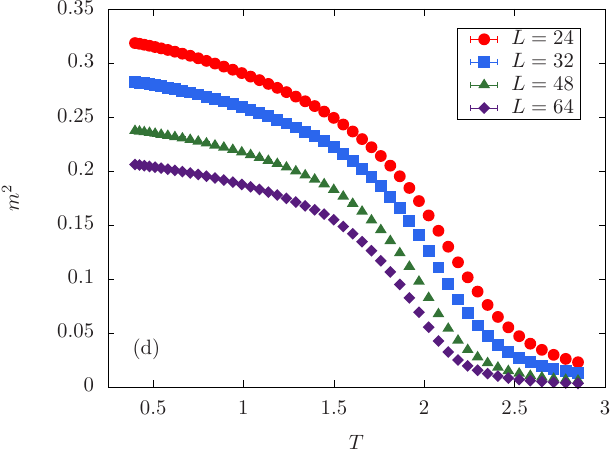}
\caption{
Binder cumulant $g_{\rm q}$ for the spin-glass order parameter as a
function of the temperature $T$ for the model described in
Ref.~\cite{lemke:96} with $\lambda = 0.50$ (a) and $\lambda = 0.75$ (b)
and system sizes $L > \ell$. In both cases the data show no crossing at
any finite temperature studied, thus suggesting that there is no
finite-temperature spin-glass phase.  Square of the magnetization $m^2$
as a function of $T$ for different system sizes for $\lambda = 0.50$ (c)
and $\lambda = 0.75$ (d).  The data decreases with increasing system
size, i.e., the system is likely in a paramagnetic phase.
\label{fig:binder}}
\end{figure*}

\section{Model and numerical details}
\label{sec:model}

In their letter \cite{lemke:96}, Lemke and Campbell argue that a
finite-temperature spin-glass transition can be induced in
two-dimensional planar topologies with next-nearest interactions. To be
precise, the model is a two-dimensional square-lattice Ising spin glass
with uniform ferromagnetic next-nearest interactions of strength $J$, in
addition to random bimodal nearest-neighbor interactions of strength
$\pm \lambda J$. The Hamiltonian of the model is
\begin{equation}
{\mathcal H} = - \sum_{\langle i,j \rangle} J_{ij} S_i S_j 
	       - J \sum_{\langle\langle i,j \rangle\rangle} S_i S_j ,
\label{eq:ham}
\end{equation}
where in Eq.~\eqref{eq:ham} $S_i \in \{\pm 1\}$ represent Ising spins on
a square lattice with $N = L^2$ sites ($L$ is the linear dimension of
the lattice).  $J = 1$ are ferromagnetic interactions between
next-nearest neighbors (denoted by $\langle\langle i,j \rangle\rangle$)
and $J_{ij} = \pm \lambda J$ are nearest-neighbor bimodally distributed
spin-glass interactions (denoted by $\langle i,j \rangle$). In our
simulations we set $J = 1$. Depending on the relative strength of the
interactions, i.e., the value of $\lambda$, Ref.~\cite{lemke:96} states
that a finite-temperature spin-glass transition can be induced in two
space dimensions. These results were further expanded in
Ref.~\cite{lemke:99}: A freezing temperature of $T_c \sim 2.1$ exists
for $\lambda = 0.5$, a ``slightly lower'' freezing temperature for
$\lambda = 0.7$, and a zero-temperature freezing for $\lambda = 1.5$. We
do emphasize that these results were produced  by relatively small
system sizes. Extensive numerical simulations by Parisi {\em et
al.}~\cite{parisi:98} find a {\em crossover} in
the critical behavior for large enough system sizes. First, from a
seemingly ordered state to a spin-glass-like state, followed by a second
crossover to a (possibly) paramagnetic state. This means that the true
thermodynamic behavior can only be observed if the system sizes exceed a
certain breakup length $\ell$.

\begin{figure}[!t]
\center
\includegraphics[width=\columnwidth]{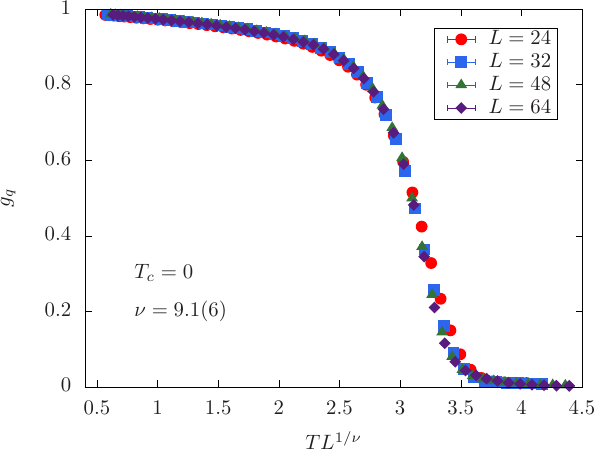}
\caption{
Finite-size scaling of the data shown in Fig.~\ref{fig:binder}(b) for
$\lambda = 0.75$ with $T_c = 0$.
\label{fig:binderscale}}
\end{figure}

However, a conclusive characterization of the critical behavior, as well
as the $\lambda$ dependence of the breakup length $\ell$ were not
discussed in detail until the extensive zero-temperature study by
Hartmann and Campbell \cite{hartmann:01d}. By computing ground-state
configurations for intermediate system sizes and estimating the
stiffness exponent that describes the scaling of energy excitations when
a domain is introduced into the system, they argue---based on {\em
zero-temperature} estimates of the spin stiffness---that there should be
a finite-temperature spin-glass transition for certain values of
$\lambda$ and linear system sizes $L$ that fulfill $L > \ell$. In
particular, they estimate that for $\lambda > \lambda_{\infty} =
0.27(8)$ no ferromagnetic order should be present. Because the breakup
length $\ell$ is large for $\lambda \sim 0.5$ ($\ell \gtrsim 45)$,
Ref.~\cite{hartmann:01d} suggests studying systems with $\lambda = 0.7$
where $\ell \approx 10$.  On the other hand, for $\lambda = 0.90$, the
stiffness exponent $\theta = 0.09(5)$ is very close to zero. Therefore,
in this work we focus on the cases where (i) we can simulate system
sizes $L \gg \ell$ and (ii) the stiffness exponent $\theta$ is clearly
positive, thus implying a finite-temperature phase, i.e., $\lambda =
0.50$ and $0.75$. A summary of the properties of the model for these
values of $\lambda$, as well as the simulation parameters are listed in
Table~\ref{tab:simparams}.  The simulations were performed using
parallel tempering Monte Carlo \cite{hukushima:96} combined with
isoenergetic cluster updates \cite{zhu:15b,houdayer:01}.  Note that we
determine the estimated value of $\theta$ for $\lambda = 0.75$ by
performing a linear fit to the data of Ref.~\cite{hartmann:01d} (quality
of fit $\sim 0.58$ \cite{press:95}) and estimate $\theta(\lambda)
\approx 1.083(3) - 1.12(4)\lambda$, valid in the window $\lambda \in
[0.5,1.1]$. Furthermore, by inspecting Fig.~7 in
Ref.~\cite{hartmann:01d}, we estimate that the breakup length for
$\lambda = 0.75$ is approximately $\ell \approx 9$.

\begin{table}
\caption{
Simulation parameters and estimates of the stiffness exponent $\theta$
and breakup length $\ell$ for different values of $\lambda$.  For both
values of $\lambda$ we studied different system sizes $L$ using parallel
tempering Monte Carlo. The lowest (highest) temperature simulated is
$T_{\rm min} = 0.4$ ($T_{\rm max} = 2.8$) with $N_T = 50$ temperature
steps. Thermalization is tested by a logarithmic binning; once the last
three bins agree within error bars we deem the system to be thermalized.
For all systems, this was the case after $N_{\rm sw} = 2^{22}$ Monte
Carlo sweeps.  Furthermore, $N_{\rm sa}$ samples were computed for each
parameter combination. Note that the estimate of $\theta$ for $\lambda 
= 0.50$ is taken from Ref.~\cite{hartmann:01d}, whereas the value for
$\lambda = 0.75$ is estimated from the published data (see text for 
details).
\label{tab:simparams}}
\begin{tabular*}{\columnwidth}{@{\extracolsep{\fill}} l l r r c c c c r}
\hline
\hline
$\lambda$ & $\theta$ & $\ell$ & $L$ & $N_{\rm sw}$ & $T_{\rm min}$ & $T_{\rm max}$ & $N_T$ & $N_{\rm sa}$ \\
\hline
$0.50$ & $0.59(8)$ & $45$ &  $48$ & $2^{22}$ & $0.4$ & $2.8$ & $50$ & $10^4$ \\
       &           &      &  $64$ & $2^{22}$ & $0.4$ & $2.8$ & $50$ & $10^4$ \\
       &           &      &  $96$ & $2^{22}$ & $0.4$ & $2.8$ & $50$ & $10^4$ \\
       &           &      & $128$ & $2^{22}$ & $0.4$ & $2.8$ & $50$ & $10^4$ \\
\hline
$0.75$ & $0.23(1)$ &  $9$ &  $24$ & $2^{22}$ & $0.4$ & $2.8$ & $50$ & $10^4$ \\
       &           &      &  $32$ & $2^{22}$ & $0.4$ & $2.8$ & $50$ & $10^4$ \\
       &           &      &  $48$ & $2^{22}$ & $0.4$ & $2.8$ & $50$ & $10^4$ \\
       &           &      &  $64$ & $2^{22}$ & $0.4$ & $2.8$ & $50$ & $10^4$ \\
\hline
\hline
\end{tabular*}
\end{table}

To detect the existence of a spin-glass transition, we measure the Binder
cumulant $g$ \cite{binder:81} of the spin-glass order parameter $q$ via
\begin{equation}
g_{\rm q} = \frac{1}{2}\left(3 - 
\frac{[\langle q^4 \rangle]_{\rm av}}{[\langle q^2 \rangle]_{\rm av}^2}\right) .
\label{eq:binder}
\end{equation}
In Eq.~\eqref{eq:binder}, $\langle \cdots \rangle$ represents a thermal
average over Monte Carlo steps and $[\cdots]_{\rm av}$ an average over
$N_{\rm sa}$ realizations of the disorder (see Table \ref{tab:simparams}
for details). The spin-glass order parameter $q$ is given by
\begin{equation}
q = \frac{1}{N} \sum_{i = 1}^N S_i^{\alpha} S_i^{\beta} ,
\label{eq:q}
\end{equation}
where ``$\alpha$'' and ``$\beta$'' represent two copies of the system
with the same disorder. The Binder cumulant is dimensionless and scales
as $g_{\rm q} = G[L^{1/\nu}(T - T_c)]$. Therefore, if $T = T_c$, data for
different system sizes cross. If, however, there is no transition, data
for different system sizes do not cross. To rule out a transition at a
temperature not simulated, a finite-size scaling of the data can be used.
Finally, we also measure the average of the square of the magnetization
$m^2 \equiv [\langle m^2 \rangle]_{\rm av}$ with
\begin{equation}
m = \frac{1}{N} \sum_{i = 1}^N S_i^{\alpha} .
\label{eq:m}
\end{equation}
Note that we measure the square of the magnetization because, on
average, $m \equiv [\langle m \rangle]_{\rm av} = 0$. Furthermore, the
magnetic susceptibility $\chi_{\rm m}$ is related to $m^2$ via
$\chi_{\rm m} = Nm^2$.

\section{Results and conclusions}
\label{sec:results}

We have performed large-scale Monte Carlo simulations of the Hamiltonian
in Eq.~\eqref{eq:ham} for system sizes $L \gg \ell$ and $\lambda = 0.50$
and $0.75$. Our results for the Binder cumulant---which should display a
crossing if there is a finite-temperature transition---are summarized in
Fig.~\ref{fig:binder}. The Binder cumulant for the spin-glass order
parameter $g_{\rm q}$ does not show a crossing down to low temperatures
for both values of $\lambda$ studied. In addition, a finite-size scaling
of the data for $\lambda = 0.75$ shown in Fig.~\ref{fig:binderscale}
strongly suggests that $T_c = 0$. Furthermore, the magnetization $m^2$
as a function of the temperature $T$ decreases with increasing system
sizes for both values of $\lambda$ studied (see Fig.~\ref{fig:binder}).
Based on these results, we conclude that the system is in a {\em
paramagnetic} state for both $\lambda = 0.50$ and $0.75$ in the
thermodynamic limit.

Our results show that the model introduced in Ref.~\cite{lemke:96} and
studied in detail in subsequent publications
\cite{parisi:98,lemke:99,hartmann:01d} does not exhibit a
finite-temperature spin-glass transition in the thermodynamic limit for
values of the parameter $\lambda$ where it is expected to show such
behavior. In agreement with the results of Ref.~\cite{parisi:98},
however for larger system sizes and and better statistics, we show that,
indeed, the thermodynamic limit is a paramagnetic phase at finite
temperature.  This also means that deducing a finite-temperature
behavior from zero-temperature simulations can be dangerous when the
system sizes are not in the thermodynamic limit \cite{hartmann:01d}.
Given recent interest in inducing finite-temperature spin-glass
transitions in quasi-planar topologies \cite{katzgraber:14}, we
conjecture that adding any set of interactions that do not grow with the
system size to a nearest-neighbor lattice will likely not result in a
finite-temperature spin-glass transition.

\begin{acknowledgments}

H.G.K.~and Z.Z.~acknowledge support from the National Science Foundation
(Grant No.~DMR-1151387). We would like to thank Texas A\&M University
for HPC resources. H.G.K.~would also like to thank Warinomo (New
Republic) for inspiration during the final stages of this manuscript.
This work is supported in part by the Office of the Director of National
Intelligence (ODNI), Intelligence Advanced Research Projects Activity
(IARPA), via MIT Lincoln Laboratory Air Force Contract
No.~FA8721-05-C-0002.  The views and conclusions contained herein are
those of the authors and should not be interpreted as necessarily
representing the official policies or endorsements, either expressed or
implied, of ODNI, IARPA, or the U.S.~Government.  The U.S.~Government is
authorized to reproduce and distribute reprints for Governmental purpose
notwithstanding any copyright annotation thereon.

\end{acknowledgments}

\bibliographystyle{apsrevtitle}
\bibliography{refs,comments}

\begin{thebibliography}{54}
\expandafter\ifx\csname natexlab\endcsname\relax\def\natexlab#1{#1}\fi
\expandafter\ifx\csname bibnamefont\endcsname\relax
  \def\bibnamefont#1{#1}\fi
\expandafter\ifx\csname bibfnamefont\endcsname\relax
  \def\bibfnamefont#1{#1}\fi
\expandafter\ifx\csname citenamefont\endcsname\relax
  \def\citenamefont#1{#1}\fi
\expandafter\ifx\csname url\endcsname\relax
  \def\url#1{\texttt{#1}}\fi
\expandafter\ifx\csname urlprefix\endcsname\relax\def\urlprefix{URL }\fi
\providecommand{\bibinfo}[2]{#2}
\providecommand{\eprint}[2][]{\url{#2}}

\bibitem[{\citenamefont{Nielsen and Chuang}(2000)}]{nielsen:00}
\bibinfo{author}{\bibfnamefont{M.~A.} \bibnamefont{Nielsen}} \bibnamefont{and}
  \bibinfo{author}{\bibfnamefont{I.~L.} \bibnamefont{Chuang}},
  \emph{\bibinfo{title}{Quantum Computation and Quantum Information}}
  (\bibinfo{publisher}{{Cambridge University Press}},
  \bibinfo{address}{Cambridge}, \bibinfo{year}{2000}).

\bibitem[{\citenamefont{Nishimori}(2001)}]{nishimori:01}
\bibinfo{author}{\bibfnamefont{H.}~\bibnamefont{Nishimori}},
  \emph{\bibinfo{title}{{Statistical Physics of Spin Glasses and Information
  Processing: An Introduction}}} (\bibinfo{publisher}{Oxford University Press},
  \bibinfo{address}{New York}, \bibinfo{year}{2001}).

\bibitem[{\citenamefont{Finnila et~al.}(1994)\citenamefont{Finnila, Gomez,
  Sebenik, Stenson, and Doll}}]{finnila:94}
\bibinfo{author}{\bibfnamefont{A.~B.} \bibnamefont{Finnila}},
  \bibinfo{author}{\bibfnamefont{M.~A.} \bibnamefont{Gomez}},
  \bibinfo{author}{\bibfnamefont{C.}~\bibnamefont{Sebenik}},
  \bibinfo{author}{\bibfnamefont{C.}~\bibnamefont{Stenson}}, \bibnamefont{and}
  \bibinfo{author}{\bibfnamefont{J.~D.} \bibnamefont{Doll}},
  \emph{\bibinfo{title}{{{Quantum annealing: A new method for minimizing
  multidimensional functions}}}}, \bibinfo{journal}{Chem. Phys. Lett.}
  \textbf{\bibinfo{volume}{219}}, \bibinfo{pages}{343} (\bibinfo{year}{1994}).

\bibitem[{\citenamefont{Kadowaki and Nishimori}(1998)}]{kadowaki:98}
\bibinfo{author}{\bibfnamefont{T.}~\bibnamefont{Kadowaki}} \bibnamefont{and}
  \bibinfo{author}{\bibfnamefont{H.}~\bibnamefont{Nishimori}},
  \emph{\bibinfo{title}{{{Quantum annealing in the transverse Ising model}}}},
  \bibinfo{journal}{Phys. Rev. E} \textbf{\bibinfo{volume}{58}},
  \bibinfo{pages}{5355} (\bibinfo{year}{1998}).

\bibitem[{\citenamefont{Brooke et~al.}(1999)\citenamefont{Brooke, Bitko,
  Rosenbaum, and Aepli}}]{brooke:99}
\bibinfo{author}{\bibfnamefont{J.}~\bibnamefont{Brooke}},
  \bibinfo{author}{\bibfnamefont{D.}~\bibnamefont{Bitko}},
  \bibinfo{author}{\bibfnamefont{T.~F.} \bibnamefont{Rosenbaum}},
  \bibnamefont{and} \bibinfo{author}{\bibfnamefont{G.}~\bibnamefont{Aepli}},
  \emph{\bibinfo{title}{Quantum annealing of a disordered magnet}},
  \bibinfo{journal}{Science} \textbf{\bibinfo{volume}{284}},
  \bibinfo{pages}{779} (\bibinfo{year}{1999}).

\bibitem[{\citenamefont{{Farhi} et~al.}(2000)\citenamefont{{Farhi},
  {Goldstone}, {Gutmann}, and {Sipser}}}]{farhi:00}
\bibinfo{author}{\bibfnamefont{E.}~\bibnamefont{{Farhi}}},
  \bibinfo{author}{\bibfnamefont{J.}~\bibnamefont{{Goldstone}}},
  \bibinfo{author}{\bibfnamefont{S.}~\bibnamefont{{Gutmann}}},
  \bibnamefont{and} \bibinfo{author}{\bibfnamefont{M.}~\bibnamefont{{Sipser}}},
  \emph{\bibinfo{title}{{{Quantum Computation by Adiabatic Evolution}}}}
  (\bibinfo{year}{2000}), \bibinfo{note}{arXiv:quant-ph/0001106}.

\bibitem[{\citenamefont{Roland and Cerf}(2002)}]{roland:02}
\bibinfo{author}{\bibfnamefont{J.}~\bibnamefont{Roland}} \bibnamefont{and}
  \bibinfo{author}{\bibfnamefont{N.~J.} \bibnamefont{Cerf}},
  \emph{\bibinfo{title}{{{Quantum search by local adiabatic evolution}}}},
  \bibinfo{journal}{Phys. Rev. A} \textbf{\bibinfo{volume}{65}},
  \bibinfo{pages}{042308} (\bibinfo{year}{2002}).

\bibitem[{\citenamefont{Santoro et~al.}(2002)\citenamefont{Santoro,
  Marto\v{n}\'ak, and Car}}]{santoro:02}
\bibinfo{author}{\bibfnamefont{G.}~\bibnamefont{Santoro}},
  \bibinfo{author}{\bibfnamefont{E.}~\bibnamefont{Marto\v{n}\'ak},
  \bibfnamefont{R.~Tosatti}}, \bibnamefont{and}
  \bibinfo{author}{\bibfnamefont{R.}~\bibnamefont{Car}},
  \emph{\bibinfo{title}{Theory of quantum annealing of an {I}sing spin glass}},
  \bibinfo{journal}{Science} \textbf{\bibinfo{volume}{295}},
  \bibinfo{pages}{2427} (\bibinfo{year}{2002}).

\bibitem[{\citenamefont{Das and Chakrabarti}(2005)}]{das:05}
\bibinfo{author}{\bibfnamefont{A.}~\bibnamefont{Das}} \bibnamefont{and}
  \bibinfo{author}{\bibfnamefont{B.~K.} \bibnamefont{Chakrabarti}},
  \emph{\bibinfo{title}{{{Quantum Annealing and Related Optimization
  Methods}}}} (\bibinfo{publisher}{Edited by A.~Das and B.K.~Chakrabarti,
  Lecture Notes in Physics 679, Berlin: Springer}, \bibinfo{year}{2005}).

\bibitem[{\citenamefont{Santoro and Tosatti}(2006)}]{santoro:06}
\bibinfo{author}{\bibfnamefont{G.~E.} \bibnamefont{Santoro}} \bibnamefont{and}
  \bibinfo{author}{\bibfnamefont{E.}~\bibnamefont{Tosatti}},
  \emph{\bibinfo{title}{{{TOPICAL REVIEW: Optimization using quantum mechanics:
  quantum annealing through adiabatic evolution}}}}, \bibinfo{journal}{J. Phys.
  A} \textbf{\bibinfo{volume}{39}}, \bibinfo{pages}{R393}
  (\bibinfo{year}{2006}).

\bibitem[{\citenamefont{Lidar}(2008)}]{lidar:08}
\bibinfo{author}{\bibfnamefont{D.~A.} \bibnamefont{Lidar}},
  \emph{\bibinfo{title}{{{Towards Fault Tolerant Adiabatic Quantum
  Computation}}}}, \bibinfo{journal}{Phys. Rev. Lett.}
  \textbf{\bibinfo{volume}{100}}, \bibinfo{pages}{160506}
  (\bibinfo{year}{2008}).

\bibitem[{\citenamefont{Das and Chakrabarti}(2008)}]{das:08}
\bibinfo{author}{\bibfnamefont{A.}~\bibnamefont{Das}} \bibnamefont{and}
  \bibinfo{author}{\bibfnamefont{B.~K.} \bibnamefont{Chakrabarti}},
  \emph{\bibinfo{title}{{{Quantum Annealing and Analog Quantum Computation}}}},
  \bibinfo{journal}{Rev. Mod. Phys.} \textbf{\bibinfo{volume}{80}},
  \bibinfo{pages}{1061} (\bibinfo{year}{2008}).

\bibitem[{\citenamefont{Morita and Nishimori}(2008)}]{morita:08}
\bibinfo{author}{\bibfnamefont{S.}~\bibnamefont{Morita}} \bibnamefont{and}
  \bibinfo{author}{\bibfnamefont{H.}~\bibnamefont{Nishimori}},
  \emph{\bibinfo{title}{{{Mathematical Foundation of Quantum Annealing}}}},
  \bibinfo{journal}{J. Math. Phys.} \textbf{\bibinfo{volume}{49}},
  \bibinfo{pages}{125210} (\bibinfo{year}{2008}).

\bibitem[{\citenamefont{Mukherjee and Chakrabarti}(2015)}]{mukherjee:15}
\bibinfo{author}{\bibfnamefont{S.}~\bibnamefont{Mukherjee}} \bibnamefont{and}
  \bibinfo{author}{\bibfnamefont{B.~K.} \bibnamefont{Chakrabarti}},
  \emph{\bibinfo{title}{{{Multivariable optimization: Quantum annealing and
  computation}}}}, \bibinfo{journal}{Eur. Phys. J. Special Topics}
  \textbf{\bibinfo{volume}{224}}, \bibinfo{pages}{17} (\bibinfo{year}{2015}).

\bibitem[{com()}]{comment:d-wave}
\urlprefix\url{http://www.dwavesys.com}.

\bibitem[{\citenamefont{{M{\' e}zard} et~al.}(1984)\citenamefont{{M{\' e}zard},
  {Parisi}, {Sourlas}, {Toulouse}, and {Virasoro}}}]{mezard:84}
\bibinfo{author}{\bibfnamefont{M.}~\bibnamefont{{M{\' e}zard}}},
  \bibinfo{author}{\bibfnamefont{G.}~\bibnamefont{{Parisi}}},
  \bibinfo{author}{\bibfnamefont{N.}~\bibnamefont{{Sourlas}}},
  \bibinfo{author}{\bibfnamefont{G.}~\bibnamefont{{Toulouse}}},
  \bibnamefont{and}
  \bibinfo{author}{\bibfnamefont{M.}~\bibnamefont{{Virasoro}}},
  \emph{\bibinfo{title}{{Nature of the Spin-Glass Phase}}},
  \bibinfo{journal}{Phys. Rev. Lett.} \textbf{\bibinfo{volume}{52}},
  \bibinfo{pages}{1156} (\bibinfo{year}{1984}).

\bibitem[{\citenamefont{Binder and Young}(1986)}]{binder:86}
\bibinfo{author}{\bibfnamefont{K.}~\bibnamefont{Binder}} \bibnamefont{and}
  \bibinfo{author}{\bibfnamefont{A.~P.} \bibnamefont{Young}},
  \emph{\bibinfo{title}{{Spin Glasses: Experimental Facts, Theoretical Concepts
  and Open Questions}}}, \bibinfo{journal}{Rev. Mod. Phys.}
  \textbf{\bibinfo{volume}{58}}, \bibinfo{pages}{801} (\bibinfo{year}{1986}).

\bibitem[{\citenamefont{Young}(1998)}]{young:98}
\bibinfo{editor}{\bibfnamefont{A.~P.} \bibnamefont{Young}}, ed.,
  \emph{\bibinfo{title}{Spin Glasses and Random Fields}}
  (\bibinfo{publisher}{World Scientific}, \bibinfo{address}{Singapore},
  \bibinfo{year}{1998}).

\bibitem[{\citenamefont{Stein and Newman}(2013)}]{stein:13}
\bibinfo{author}{\bibfnamefont{D.~L.} \bibnamefont{Stein}} \bibnamefont{and}
  \bibinfo{author}{\bibfnamefont{C.~M.} \bibnamefont{Newman}},
  \emph{\bibinfo{title}{{Spin Glasses and Complexity}}}, Primers in Complex
  Systems (\bibinfo{publisher}{Princeton University Press},
  \bibinfo{address}{Princeton NJ}, \bibinfo{year}{2013}).

\bibitem[{\citenamefont{{Dickson} et~al.}(2013)\citenamefont{{Dickson},
  {Johnson}, {Amin}, {Harris}, {Altomare}, {Berkley}, {Bunyk}, {Cai},
  {Chapple}, {Chavez} et~al.}}]{dickson:13}
\bibinfo{author}{\bibfnamefont{N.~G.} \bibnamefont{{Dickson}}},
  \bibinfo{author}{\bibfnamefont{M.~W.} \bibnamefont{{Johnson}}},
  \bibinfo{author}{\bibfnamefont{M.~H.} \bibnamefont{{Amin}}},
  \bibinfo{author}{\bibfnamefont{R.}~\bibnamefont{{Harris}}},
  \bibinfo{author}{\bibfnamefont{F.}~\bibnamefont{{Altomare}}},
  \bibinfo{author}{\bibfnamefont{A.~J.} \bibnamefont{{Berkley}}},
  \bibinfo{author}{\bibfnamefont{P.}~\bibnamefont{{Bunyk}}},
  \bibinfo{author}{\bibfnamefont{J.}~\bibnamefont{{Cai}}},
  \bibinfo{author}{\bibfnamefont{E.~M.} \bibnamefont{{Chapple}}},
  \bibinfo{author}{\bibfnamefont{P.}~\bibnamefont{{Chavez}}},
  \bibnamefont{et~al.}, \emph{\bibinfo{title}{{{Thermally assisted quantum
  annealing of a 16-qubit problem}}}}, \bibinfo{journal}{Nat. Commun.}
  \textbf{\bibinfo{volume}{4}}, \bibinfo{pages}{1903} (\bibinfo{year}{2013}).

\bibitem[{\citenamefont{Pudenz et~al.}(2014)\citenamefont{Pudenz, Albash, and
  Lidar}}]{pudenz:13}
\bibinfo{author}{\bibfnamefont{K.~L.} \bibnamefont{Pudenz}},
  \bibinfo{author}{\bibfnamefont{T.}~\bibnamefont{Albash}}, \bibnamefont{and}
  \bibinfo{author}{\bibfnamefont{D.~A.} \bibnamefont{Lidar}},
  \emph{\bibinfo{title}{Error-corrected quantum annealing with hundreds of
  qubits}}, \bibinfo{journal}{Nat. Commun.} \textbf{\bibinfo{volume}{5}},
  \bibinfo{pages}{3243} (\bibinfo{year}{2014}).

\bibitem[{\citenamefont{Smith and Smolin}(2013)}]{smith:13}
\bibinfo{author}{\bibfnamefont{G.}~\bibnamefont{Smith}} \bibnamefont{and}
  \bibinfo{author}{\bibfnamefont{J.}~\bibnamefont{Smolin}},
  \emph{\bibinfo{title}{{{Putting ``Quantumness'' to the Test}}}},
  \bibinfo{journal}{Physics} \textbf{\bibinfo{volume}{6}}, \bibinfo{pages}{105}
  (\bibinfo{year}{2013}).

\bibitem[{\citenamefont{{Boixo} et~al.}(2013)\citenamefont{{Boixo}, {Albash},
  {Spedalieri}, {Chancellor}, and {Lidar}}}]{boixo:13a}
\bibinfo{author}{\bibfnamefont{S.}~\bibnamefont{{Boixo}}},
  \bibinfo{author}{\bibfnamefont{T.}~\bibnamefont{{Albash}}},
  \bibinfo{author}{\bibfnamefont{F.~M.} \bibnamefont{{Spedalieri}}},
  \bibinfo{author}{\bibfnamefont{N.}~\bibnamefont{{Chancellor}}},
  \bibnamefont{and} \bibinfo{author}{\bibfnamefont{D.~A.}
  \bibnamefont{{Lidar}}}, \emph{\bibinfo{title}{{{Experimental signature of
  programmable quantum annealing}}}}, \bibinfo{journal}{Nat. Commun.}
  \textbf{\bibinfo{volume}{4}}, \bibinfo{pages}{2067} (\bibinfo{year}{2013}).

\bibitem[{\citenamefont{Albash et~al.}(2015{\natexlab{a}})\citenamefont{Albash,
  R{\o}nnow, Troyer, and Lidar}}]{albash:15a}
\bibinfo{author}{\bibfnamefont{T.}~\bibnamefont{Albash}},
  \bibinfo{author}{\bibfnamefont{T.~F.} \bibnamefont{R{\o}nnow}},
  \bibinfo{author}{\bibfnamefont{M.}~\bibnamefont{Troyer}}, \bibnamefont{and}
  \bibinfo{author}{\bibfnamefont{D.~A.} \bibnamefont{Lidar}},
  \emph{\bibinfo{title}{{{Reexamining classical and quantum models for the
  D-Wave One processor}}}}, \bibinfo{journal}{Eur. Phys. J. Spec. Top.}
  \textbf{\bibinfo{volume}{224}}, \bibinfo{pages}{111}
  (\bibinfo{year}{2015}{\natexlab{a}}).

\bibitem[{\citenamefont{{R{\o}nnow} et~al.}(2014)\citenamefont{{R{\o}nnow},
  {Wang}, {Job}, {Boixo}, {Isakov}, {Wecker}, {Martinis}, {Lidar}, and
  {Troyer}}}]{ronnow:14a}
\bibinfo{author}{\bibfnamefont{T.~F.} \bibnamefont{{R{\o}nnow}}},
  \bibinfo{author}{\bibfnamefont{Z.}~\bibnamefont{{Wang}}},
  \bibinfo{author}{\bibfnamefont{J.}~\bibnamefont{{Job}}},
  \bibinfo{author}{\bibfnamefont{S.}~\bibnamefont{{Boixo}}},
  \bibinfo{author}{\bibfnamefont{S.~V.} \bibnamefont{{Isakov}}},
  \bibinfo{author}{\bibfnamefont{D.}~\bibnamefont{{Wecker}}},
  \bibinfo{author}{\bibfnamefont{J.~M.} \bibnamefont{{Martinis}}},
  \bibinfo{author}{\bibfnamefont{D.~A.} \bibnamefont{{Lidar}}},
  \bibnamefont{and} \bibinfo{author}{\bibfnamefont{M.}~\bibnamefont{{Troyer}}},
  \emph{\bibinfo{title}{{Defining and detecting quantum speedup}}},
  \bibinfo{journal}{Science} \textbf{\bibinfo{volume}{345}},
  \bibinfo{pages}{420} (\bibinfo{year}{2014}).

\bibitem[{\citenamefont{Katzgraber et~al.}(2014)\citenamefont{Katzgraber,
  Hamze, and Andrist}}]{katzgraber:14}
\bibinfo{author}{\bibfnamefont{H.~G.} \bibnamefont{Katzgraber}},
  \bibinfo{author}{\bibfnamefont{F.}~\bibnamefont{Hamze}}, \bibnamefont{and}
  \bibinfo{author}{\bibfnamefont{R.~S.} \bibnamefont{Andrist}},
  \emph{\bibinfo{title}{{Glassy Chimeras Could Be Blind to Quantum Speedup:
  Designing Better Benchmarks for Quantum Annealing Machines}}},
  \bibinfo{journal}{Phys. Rev. X} \textbf{\bibinfo{volume}{4}},
  \bibinfo{pages}{021008} (\bibinfo{year}{2014}).

\bibitem[{\citenamefont{{Lanting} et~al.}(2014)\citenamefont{{Lanting},
  {Przybysz}, {Smirnov}, {Spedalieri}, {Amin}, {Berkley}, {Harris}, {Altomare},
  {Boixo}, {Bunyk} et~al.}}]{lanting:14}
\bibinfo{author}{\bibfnamefont{T.}~\bibnamefont{{Lanting}}},
  \bibinfo{author}{\bibfnamefont{A.~J.} \bibnamefont{{Przybysz}}},
  \bibinfo{author}{\bibfnamefont{A.~Y.} \bibnamefont{{Smirnov}}},
  \bibinfo{author}{\bibfnamefont{F.~M.} \bibnamefont{{Spedalieri}}},
  \bibinfo{author}{\bibfnamefont{M.~H.} \bibnamefont{{Amin}}},
  \bibinfo{author}{\bibfnamefont{A.~J.} \bibnamefont{{Berkley}}},
  \bibinfo{author}{\bibfnamefont{R.}~\bibnamefont{{Harris}}},
  \bibinfo{author}{\bibfnamefont{F.}~\bibnamefont{{Altomare}}},
  \bibinfo{author}{\bibfnamefont{S.}~\bibnamefont{{Boixo}}},
  \bibinfo{author}{\bibfnamefont{P.}~\bibnamefont{{Bunyk}}},
  \bibnamefont{et~al.}, \emph{\bibinfo{title}{Entanglement in a quantum
  annealing processor}}, \bibinfo{journal}{Phys. Rev. X}
  \textbf{\bibinfo{volume}{4}}, \bibinfo{pages}{021041} (\bibinfo{year}{2014}).

\bibitem[{\citenamefont{{Santra} et~al.}(2014)\citenamefont{{Santra}, {Quiroz},
  {Ver Steeg}, and {Lidar}}}]{santra:14}
\bibinfo{author}{\bibfnamefont{S.}~\bibnamefont{{Santra}}},
  \bibinfo{author}{\bibfnamefont{G.}~\bibnamefont{{Quiroz}}},
  \bibinfo{author}{\bibfnamefont{G.}~\bibnamefont{{Ver Steeg}}},
  \bibnamefont{and} \bibinfo{author}{\bibfnamefont{D.~A.}
  \bibnamefont{{Lidar}}}, \emph{\bibinfo{title}{{{Max 2-SAT with up to 108
  qubits}}}}, \bibinfo{journal}{New J. Phys.} \textbf{\bibinfo{volume}{16}},
  \bibinfo{pages}{045006} (\bibinfo{year}{2014}).

\bibitem[{\citenamefont{Shin et~al.}(2014)\citenamefont{Shin, Smith, Smolin,
  and Vazirani}}]{shin:14}
\bibinfo{author}{\bibfnamefont{S.~W.} \bibnamefont{Shin}},
  \bibinfo{author}{\bibfnamefont{G.}~\bibnamefont{Smith}},
  \bibinfo{author}{\bibfnamefont{J.~A.} \bibnamefont{Smolin}},
  \bibnamefont{and} \bibinfo{author}{\bibfnamefont{U.}~\bibnamefont{Vazirani}},
  \emph{\bibinfo{title}{{{How ``Quantum'' is the D-Wave Machine?}}}}
  (\bibinfo{year}{2014}), \bibinfo{note}{(arXiv:1401.7087)}.

\bibitem[{\citenamefont{{Boixo} et~al.}(2014)\citenamefont{{Boixo},
  {R{\o}nnow}, {Isakov}, {Wang}, {Wecker}, {Lidar}, {Martinis}, and
  {Troyer}}}]{boixo:14}
\bibinfo{author}{\bibfnamefont{S.}~\bibnamefont{{Boixo}}},
  \bibinfo{author}{\bibfnamefont{T.~F.} \bibnamefont{{R{\o}nnow}}},
  \bibinfo{author}{\bibfnamefont{S.~V.} \bibnamefont{{Isakov}}},
  \bibinfo{author}{\bibfnamefont{Z.}~\bibnamefont{{Wang}}},
  \bibinfo{author}{\bibfnamefont{D.}~\bibnamefont{{Wecker}}},
  \bibinfo{author}{\bibfnamefont{D.~A.} \bibnamefont{{Lidar}}},
  \bibinfo{author}{\bibfnamefont{J.~M.} \bibnamefont{{Martinis}}},
  \bibnamefont{and} \bibinfo{author}{\bibfnamefont{M.}~\bibnamefont{{Troyer}}},
  \emph{\bibinfo{title}{{Evidence for quantum annealing with more than one
  hundred qubits}}}, \bibinfo{journal}{Nat. Phys.}
  \textbf{\bibinfo{volume}{10}}, \bibinfo{pages}{218} (\bibinfo{year}{2014}).

\bibitem[{\citenamefont{Albash et~al.}(2015{\natexlab{b}})\citenamefont{Albash,
  Vinci, Mishra, Warburton, and Lidar}}]{albash:15}
\bibinfo{author}{\bibfnamefont{T.}~\bibnamefont{Albash}},
  \bibinfo{author}{\bibfnamefont{W.}~\bibnamefont{Vinci}},
  \bibinfo{author}{\bibfnamefont{A.}~\bibnamefont{Mishra}},
  \bibinfo{author}{\bibfnamefont{P.~A.} \bibnamefont{Warburton}},
  \bibnamefont{and} \bibinfo{author}{\bibfnamefont{D.~A.} \bibnamefont{Lidar}},
  \emph{\bibinfo{title}{{{Consistency Tests of Classical and Quantum Models for
  a Quantum Device}}}}, \bibinfo{journal}{Phys. Rev. A}
  \textbf{\bibinfo{volume}{91}}, \bibinfo{pages}{042314}
  (\bibinfo{year}{2015}{\natexlab{b}}).

\bibitem[{\citenamefont{Katzgraber et~al.}(2015)\citenamefont{Katzgraber,
  Hamze, Zhu, Ochoa, and Munoz-Bauza}}]{katzgraber:15}
\bibinfo{author}{\bibfnamefont{H.~G.} \bibnamefont{Katzgraber}},
  \bibinfo{author}{\bibfnamefont{F.}~\bibnamefont{Hamze}},
  \bibinfo{author}{\bibfnamefont{Z.}~\bibnamefont{Zhu}},
  \bibinfo{author}{\bibfnamefont{A.~J.} \bibnamefont{Ochoa}}, \bibnamefont{and}
  \bibinfo{author}{\bibfnamefont{H.}~\bibnamefont{Munoz-Bauza}},
  \emph{\bibinfo{title}{{Seeking Quantum Speedup Through Spin Glasses: The
  Good, the Bad, and the Ugly}}}, \bibinfo{journal}{Phys. Rev. X}
  \textbf{\bibinfo{volume}{5}}, \bibinfo{pages}{031026} (\bibinfo{year}{2015}).

\bibitem[{\citenamefont{{Martin-Mayor} and {Hen}}(2015)}]{martin-mayor:15a}
\bibinfo{author}{\bibfnamefont{V.}~\bibnamefont{{Martin-Mayor}}}
  \bibnamefont{and} \bibinfo{author}{\bibfnamefont{I.}~\bibnamefont{{Hen}}},
  \emph{\bibinfo{title}{{{Unraveling Quantum Annealers using Classical
  Hardness}}}}, \bibinfo{journal}{Nature Scientific Reports}
  \textbf{\bibinfo{volume}{5}}, \bibinfo{pages}{15324} (\bibinfo{year}{2015}).

\bibitem[{\citenamefont{Pudenz et~al.}(2015)\citenamefont{Pudenz, Albash, and
  Lidar}}]{pudenz:15}
\bibinfo{author}{\bibfnamefont{K.~L.} \bibnamefont{Pudenz}},
  \bibinfo{author}{\bibfnamefont{T.}~\bibnamefont{Albash}}, \bibnamefont{and}
  \bibinfo{author}{\bibfnamefont{D.~A.} \bibnamefont{Lidar}},
  \emph{\bibinfo{title}{{Quantum Annealing Correction for Random Ising
  Problems}}}, \bibinfo{journal}{Phys. Rev. A} \textbf{\bibinfo{volume}{91}},
  \bibinfo{pages}{042302} (\bibinfo{year}{2015}).

\bibitem[{\citenamefont{Hen et~al.}(2015)\citenamefont{Hen, Job, Albash,
  R{\o}nnow, Troyer, and Lidar}}]{hen:15a}
\bibinfo{author}{\bibfnamefont{I.}~\bibnamefont{Hen}},
  \bibinfo{author}{\bibfnamefont{J.}~\bibnamefont{Job}},
  \bibinfo{author}{\bibfnamefont{T.}~\bibnamefont{Albash}},
  \bibinfo{author}{\bibfnamefont{T.~F.} \bibnamefont{R{\o}nnow}},
  \bibinfo{author}{\bibfnamefont{M.}~\bibnamefont{Troyer}}, \bibnamefont{and}
  \bibinfo{author}{\bibfnamefont{D.~A.} \bibnamefont{Lidar}},
  \emph{\bibinfo{title}{{Probing for quantum speedup in spin-glass problems
  with planted solutions}}}, \bibinfo{journal}{Phys. Rev. A}
  \textbf{\bibinfo{volume}{92}}, \bibinfo{pages}{042325}
  (\bibinfo{year}{2015}).

\bibitem[{\citenamefont{Venturelli et~al.}(2015)\citenamefont{Venturelli,
  Mandr{\`a}, Knysh, O'Gorman, Biswas, and Smelyanskiy}}]{venturelli:15a}
\bibinfo{author}{\bibfnamefont{D.}~\bibnamefont{Venturelli}},
  \bibinfo{author}{\bibfnamefont{S.}~\bibnamefont{Mandr{\`a}}},
  \bibinfo{author}{\bibfnamefont{S.}~\bibnamefont{Knysh}},
  \bibinfo{author}{\bibfnamefont{B.}~\bibnamefont{O'Gorman}},
  \bibinfo{author}{\bibfnamefont{R.}~\bibnamefont{Biswas}}, \bibnamefont{and}
  \bibinfo{author}{\bibfnamefont{V.}~\bibnamefont{Smelyanskiy}},
  \emph{\bibinfo{title}{{Quantum Optimization of Fully Connected Spin
  Glasses}}}, \bibinfo{journal}{Phys. Rev. X} \textbf{\bibinfo{volume}{5}},
  \bibinfo{pages}{031040} (\bibinfo{year}{2015}).

\bibitem[{\citenamefont{Vinci et~al.}(2015)\citenamefont{Vinci, Albash,
  Paz-Silva, Hen, and Lidar}}]{vinci:15}
\bibinfo{author}{\bibfnamefont{W.}~\bibnamefont{Vinci}},
  \bibinfo{author}{\bibfnamefont{T.}~\bibnamefont{Albash}},
  \bibinfo{author}{\bibfnamefont{G.}~\bibnamefont{Paz-Silva}},
  \bibinfo{author}{\bibfnamefont{I.}~\bibnamefont{Hen}}, \bibnamefont{and}
  \bibinfo{author}{\bibfnamefont{D.~A.} \bibnamefont{Lidar}},
  \emph{\bibinfo{title}{Quantum annealing correction with minor embedding}},
  \bibinfo{journal}{Phys. Rev. A} \textbf{\bibinfo{volume}{92}},
  \bibinfo{pages}{042310} (\bibinfo{year}{2015}).

\bibitem[{\citenamefont{Zhu et~al.}(2016)\citenamefont{Zhu, Ochoa, Hamze,
  Schnabel, and Katzgraber}}]{zhu:16}
\bibinfo{author}{\bibfnamefont{Z.}~\bibnamefont{Zhu}},
  \bibinfo{author}{\bibfnamefont{A.~J.} \bibnamefont{Ochoa}},
  \bibinfo{author}{\bibfnamefont{F.}~\bibnamefont{Hamze}},
  \bibinfo{author}{\bibfnamefont{S.}~\bibnamefont{Schnabel}}, \bibnamefont{and}
  \bibinfo{author}{\bibfnamefont{H.~G.} \bibnamefont{Katzgraber}},
  \emph{\bibinfo{title}{{{Best-case performance of quantum annealers on native
  spin-glass benchmarks: How chaos can affect success probabilities}}}},
  \bibinfo{journal}{Phys. Rev. A} \textbf{\bibinfo{volume}{93}},
  \bibinfo{pages}{012317} (\bibinfo{year}{2016}).

\bibitem[{\citenamefont{Denchev et~al.}(2016)\citenamefont{Denchev, Boixo,
  Isakov, Ding, Babbush, Smelyanskiy, Martinis, and Neven}}]{denchev:16}
\bibinfo{author}{\bibfnamefont{V.~S.} \bibnamefont{Denchev}},
  \bibinfo{author}{\bibfnamefont{S.}~\bibnamefont{Boixo}},
  \bibinfo{author}{\bibfnamefont{S.~V.} \bibnamefont{Isakov}},
  \bibinfo{author}{\bibfnamefont{N.}~\bibnamefont{Ding}},
  \bibinfo{author}{\bibfnamefont{R.}~\bibnamefont{Babbush}},
  \bibinfo{author}{\bibfnamefont{V.}~\bibnamefont{Smelyanskiy}},
  \bibinfo{author}{\bibfnamefont{J.}~\bibnamefont{Martinis}}, \bibnamefont{and}
  \bibinfo{author}{\bibfnamefont{H.}~\bibnamefont{Neven}},
  \emph{\bibinfo{title}{{W}hat is the {C}omputational {V}alue of {F}inite
  {R}ange {T}unneling?}}, \bibinfo{journal}{Phys. Rev. X}
  \textbf{\bibinfo{volume}{6}}, \bibinfo{pages}{031015} (\bibinfo{year}{2016}).

\bibitem[{\citenamefont{Bunyk et~al.}(2014)\citenamefont{Bunyk, Hoskinson,
  Johnson, Tolkacheva, Altomare, Berkley, Harris, Hilton, Lanting, and
  Whittaker}}]{bunyk:14}
\bibinfo{author}{\bibfnamefont{P.}~\bibnamefont{Bunyk}},
  \bibinfo{author}{\bibfnamefont{E.}~\bibnamefont{Hoskinson}},
  \bibinfo{author}{\bibfnamefont{M.~W.} \bibnamefont{Johnson}},
  \bibinfo{author}{\bibfnamefont{E.}~\bibnamefont{Tolkacheva}},
  \bibinfo{author}{\bibfnamefont{F.}~\bibnamefont{Altomare}},
  \bibinfo{author}{\bibfnamefont{A.~J.} \bibnamefont{Berkley}},
  \bibinfo{author}{\bibfnamefont{R.}~\bibnamefont{Harris}},
  \bibinfo{author}{\bibfnamefont{J.~P.} \bibnamefont{Hilton}},
  \bibinfo{author}{\bibfnamefont{T.}~\bibnamefont{Lanting}}, \bibnamefont{and}
  \bibinfo{author}{\bibfnamefont{J.}~\bibnamefont{Whittaker}},
  \emph{\bibinfo{title}{{Architectural Considerations in the Design of a
  Superconducting Quantum Annealing Processor}}}, \bibinfo{journal}{IEEE Trans.
  Appl. Supercond.} \textbf{\bibinfo{volume}{24}}, \bibinfo{pages}{1}
  (\bibinfo{year}{2014}).

\bibitem[{\citenamefont{Kirkpatrick et~al.}(1983)\citenamefont{Kirkpatrick,
  {Gelatt, Jr.}, and Vecchi}}]{kirkpatrick:83}
\bibinfo{author}{\bibfnamefont{S.}~\bibnamefont{Kirkpatrick}},
  \bibinfo{author}{\bibfnamefont{C.~D.} \bibnamefont{{Gelatt, Jr.}}},
  \bibnamefont{and} \bibinfo{author}{\bibfnamefont{M.~P.}
  \bibnamefont{Vecchi}}, \emph{\bibinfo{title}{Optimization by simulated
  annealing}}, \bibinfo{journal}{Science} \textbf{\bibinfo{volume}{220}},
  \bibinfo{pages}{671} (\bibinfo{year}{1983}).

\bibitem[{\citenamefont{{Mandr{\`a}} et~al.}(2016)\citenamefont{{Mandr{\`a}},
  {Zhu}, {Wang}, {Perdomo-Ortiz}, and {Katzgraber}}}]{mandra:16b}
\bibinfo{author}{\bibfnamefont{S.}~\bibnamefont{{Mandr{\`a}}}},
  \bibinfo{author}{\bibfnamefont{Z.}~\bibnamefont{{Zhu}}},
  \bibinfo{author}{\bibfnamefont{W.}~\bibnamefont{{Wang}}},
  \bibinfo{author}{\bibfnamefont{A.}~\bibnamefont{{Perdomo-Ortiz}}},
  \bibnamefont{and} \bibinfo{author}{\bibfnamefont{H.~G.}
  \bibnamefont{{Katzgraber}}}, \emph{\bibinfo{title}{{Strengths and weaknesses
  of weak-strong cluster problems: A detailed overview of state-of-the-art
  classical heuristics versus quantum approaches}}}, \bibinfo{journal}{Phys.
  Rev. A} \textbf{\bibinfo{volume}{94}}, \bibinfo{pages}{022337}
  (\bibinfo{year}{2016}).

\bibitem[{\citenamefont{{Lemke} and {Campbell}}(1996)}]{lemke:96}
\bibinfo{author}{\bibfnamefont{N.}~\bibnamefont{{Lemke}}} \bibnamefont{and}
  \bibinfo{author}{\bibfnamefont{I.~A.} \bibnamefont{{Campbell}}},
  \emph{\bibinfo{title}{{{Two-Dimensional Ising Spin Glasses with Nonzero
  Ordering Temperatures}}}}, \bibinfo{journal}{Phys. Rev. Lett.}
  \textbf{\bibinfo{volume}{76}}, \bibinfo{pages}{4616} (\bibinfo{year}{1996}).

\bibitem[{\citenamefont{{Parisi} et~al.}(1998)\citenamefont{{Parisi},
  {Ruiz-Lorenzo}, and {Stariolo}}}]{parisi:98}
\bibinfo{author}{\bibfnamefont{G.}~\bibnamefont{{Parisi}}},
  \bibinfo{author}{\bibfnamefont{J.~J.} \bibnamefont{{Ruiz-Lorenzo}}},
  \bibnamefont{and} \bibinfo{author}{\bibfnamefont{D.~A.}
  \bibnamefont{{Stariolo}}}, \emph{\bibinfo{title}{{{Crossovers in the
  two-dimensional Ising spin glass with ferromagnetic next-nearest-neighbour
  interactions}}}}, \bibinfo{journal}{J. Phys. A}
  \textbf{\bibinfo{volume}{31}}, \bibinfo{pages}{4657} (\bibinfo{year}{1998}).

\bibitem[{\citenamefont{{Lemke} and {Campbell}}(1999)}]{lemke:99}
\bibinfo{author}{\bibfnamefont{N.}~\bibnamefont{{Lemke}}} \bibnamefont{and}
  \bibinfo{author}{\bibfnamefont{I.~A.} \bibnamefont{{Campbell}}},
  \emph{\bibinfo{title}{{Finite-temperature phase transition in the
  two-dimensional randomly coupled ferromagnet}}}, \bibinfo{journal}{J. Phys.
  A} \textbf{\bibinfo{volume}{32}}, \bibinfo{pages}{7851}
  (\bibinfo{year}{1999}).

\bibitem[{\citenamefont{{Hartmann} and {Campbell}}(2001)}]{hartmann:01d}
\bibinfo{author}{\bibfnamefont{A.~K.} \bibnamefont{{Hartmann}}}
  \bibnamefont{and} \bibinfo{author}{\bibfnamefont{I.~A.}
  \bibnamefont{{Campbell}}}, \emph{\bibinfo{title}{{Ordered phase in the
  two-dimensional randomly coupled ferromagnet}}}, \bibinfo{journal}{Phys. Rev.
  B} \textbf{\bibinfo{volume}{63}}, \bibinfo{pages}{094423}
  (\bibinfo{year}{2001}).

\bibitem[{\citenamefont{Bray and Moore}(1984)}]{bray:84}
\bibinfo{author}{\bibfnamefont{A.~J.} \bibnamefont{Bray}} \bibnamefont{and}
  \bibinfo{author}{\bibfnamefont{M.~A.} \bibnamefont{Moore}},
  \emph{\bibinfo{title}{Lower critical dimension of {I}sing spin glasses: a
  numerical study}}, \bibinfo{journal}{J. Phys. C}
  \textbf{\bibinfo{volume}{17}}, \bibinfo{pages}{L463} (\bibinfo{year}{1984}).

\bibitem[{\citenamefont{Hartmann and Rieger}(2001)}]{hartmann:01}
\bibinfo{author}{\bibfnamefont{A.~K.} \bibnamefont{Hartmann}} \bibnamefont{and}
  \bibinfo{author}{\bibfnamefont{H.}~\bibnamefont{Rieger}},
  \emph{\bibinfo{title}{Optimization Algorithms in Physics}}
  (\bibinfo{publisher}{Wiley-VCH}, \bibinfo{address}{Berlin},
  \bibinfo{year}{2001}).

\bibitem[{\citenamefont{Boettcher}(2005)}]{boettcher:05d}
\bibinfo{author}{\bibfnamefont{S.}~\bibnamefont{Boettcher}},
  \emph{\bibinfo{title}{{Stiffness of the {E}dwards-{A}nderson Model in all
  Dimensions}}}, \bibinfo{journal}{Phys. Rev. Lett.}
  \textbf{\bibinfo{volume}{95}}, \bibinfo{pages}{197205}
  (\bibinfo{year}{2005}).

\bibitem[{\citenamefont{Hukushima and Nemoto}(1996)}]{hukushima:96}
\bibinfo{author}{\bibfnamefont{K.}~\bibnamefont{Hukushima}} \bibnamefont{and}
  \bibinfo{author}{\bibfnamefont{K.}~\bibnamefont{Nemoto}},
  \emph{\bibinfo{title}{Exchange {M}onte {C}arlo method and application to spin
  glass simulations}}, \bibinfo{journal}{J. Phys. Soc. Jpn.}
  \textbf{\bibinfo{volume}{65}}, \bibinfo{pages}{1604} (\bibinfo{year}{1996}).

\bibitem[{\citenamefont{{Zhu} et~al.}(2015)\citenamefont{{Zhu}, {Ochoa}, and
  {Katzgraber}}}]{zhu:15b}
\bibinfo{author}{\bibfnamefont{Z.}~\bibnamefont{{Zhu}}},
  \bibinfo{author}{\bibfnamefont{A.~J.} \bibnamefont{{Ochoa}}},
  \bibnamefont{and} \bibinfo{author}{\bibfnamefont{H.~G.}
  \bibnamefont{{Katzgraber}}}, \emph{\bibinfo{title}{{{Efficient Cluster
  Algorithm for Spin Glasses in Any Space Dimension}}}},
  \bibinfo{journal}{Phys. Rev. Lett.} \textbf{\bibinfo{volume}{115}},
  \bibinfo{pages}{077201} (\bibinfo{year}{2015}).

\bibitem[{\citenamefont{Houdayer}(2001)}]{houdayer:01}
\bibinfo{author}{\bibfnamefont{J.}~\bibnamefont{Houdayer}},
  \emph{\bibinfo{title}{A cluster {M}onte {C}arlo algorithm for 2-dimensional
  spin glasses}}, \bibinfo{journal}{Eur. Phys. J. B.}
  \textbf{\bibinfo{volume}{22}}, \bibinfo{pages}{479} (\bibinfo{year}{2001}).

\bibitem[{\citenamefont{Press et~al.}(1995)\citenamefont{Press, Teukolsky,
  Vetterling, and Flannery}}]{press:95}
\bibinfo{author}{\bibfnamefont{W.~H.} \bibnamefont{Press}},
  \bibinfo{author}{\bibfnamefont{S.~A.} \bibnamefont{Teukolsky}},
  \bibinfo{author}{\bibfnamefont{W.~T.} \bibnamefont{Vetterling}},
  \bibnamefont{and} \bibinfo{author}{\bibfnamefont{B.~P.}
  \bibnamefont{Flannery}}, \emph{\bibinfo{title}{Numerical Recipes in C}}
  (\bibinfo{publisher}{Cambridge University Press},
  \bibinfo{address}{Cambridge, England}, \bibinfo{year}{1995}).

\bibitem[{\citenamefont{Binder}(1981)}]{binder:81}
\bibinfo{author}{\bibfnamefont{K.}~\bibnamefont{Binder}},
  \emph{\bibinfo{title}{Critical properties from {M}onte {C}arlo coarse
  graining and renormalization}}, \bibinfo{journal}{Phys. Rev. Lett.}
  \textbf{\bibinfo{volume}{47}}, \bibinfo{pages}{693} (\bibinfo{year}{1981}).

\end{thebibliography}

\end{document}